# A 2D T-carbon 2-(111) structure with tunable electric and optical properties via chemical decorations: a first-principles investigation

Haifang Cai [a], Zhiwen Duan [b], Douglas S. Galvao [c*], and Kun Cai [b*]

[a] School of Civil Engineering, Yan' an University, Yan' an 716000, China
[b] School of Science, Harbin Institute of Technology, Shenzhen 518055, China
[c] Applied Physics Department and Center for Computational Engineering & Sciences - CCES, State University of Campinas, Campinas, SP 13081-970, Brazil.
*Corresponding authors' email addresses: kun.cai@hit.edu.cn (K.C.); galvao@ifi.unicamp.br (D.S.G.)

## Abstract

We proposed a new two-dimensional carbon material named 2-(111) planar T-carbon, which is obtained by slicing bulk T-carbon along its (111) crystallographic direction. 2-(111) planar T-carbon's optical and electrical properties can be engineered via surface decoration. Comparing the DFT phonon spectra of pristine and five decorated 2-(111) planar T-carbon obtained by first-principles calculations, we conclude that surface decoration presents a promising, effective, and feasible strategy to improve the structural stability of 2-(111) planar T-carbon. The calculated band structures and electronic properties show direct electronic band gap values between 0.17 eV (-O= decorated) and 2.21 eV (Hydrogenated). Chemical decoration also promises blue or red energy shifts in its optical properties.

**Keywords:** 2-(111) planar T-carbon, surface decoration, electronic properties, optical properties, first-principles calculation.

# 1. Introduction

In the periodic table, carbon is one of the most fundamental elements. The allotropes of carbon have been recognized and utilized since ancient times, and a series of its compounds are referred to as organic substances, considered fundamental to life. Carbon has numerous allotropes, including graphite, c-diamond, h-diamond[1], amorphous carbon, fullerene[2], carbon nanotubes[3], and graphene[4]. Natural graphite and diamond can be found in nature, while other allotropes are typically synthesized in laboratories. The *sp*$^2$-type carbon allotropes, i.e., fullerene, carbon nanotubes, and graphene, have significantly impacted fields such as chemistry, physics, materials science, and information technology. More methods have been continuously developed to synthesize and predict new carbon allotropes, mainly through first-principles calculations, which are crucial for discovering new allotropes[5-13]. These calculations enable the prediction of the existence and properties of carbon allotropes such as M-carbon[10], bct-C4[11], W-carbon[12], Y-carbon, and TY-carbon[13]. These materials exhibit many excellent physical properties and are widely used in electronics, optics, catalysis, adsorption, and energy fields. The unique properties of these carbon allotropes have attracted increasing attention from researchers, and according to the Samara Carbon Allotrope Database (SACADA), over 500 types of carbon allotropes have been proposed[14].

Based on first-principles simulations, in 2011, Sheng et al. proposed a new carbon allotrope named T-carbon[15]. It was created by replacing each carbon atom in the diamond lattice structure with a carbon tetrahedron. T-carbon has the same space group (*Fd*3m) as diamond. To determine its structural stability, they calculated its phonon spectrum and verified the absence of negative frequencies.

The obtained optimized T-carbon lattice constant is 7.52 Å[15], more than twice that of diamond (~3.566 Å). Unlike diamond, which has just a single type of C-C bond (~1.544 Å), T-carbon has two: intra-tetrahedral (~1.502 Å) and inter-tetrahedral bonds (~1.417 Å). T-carbon has a relatively low density (~1.50 g/cm³), much lower than diamond and 32% less than that of graphite. It also possesses a direct electronic bandgap of about 3.0 eV, making it a semiconductor[15]. As a structure, T-carbon is relatively "fluffy," with large ring gaps formed between tetrahedrons, suggesting its potential use as a hydrogen storage material[16] (see **Fig. 1**).

In 2017, Zhang and Su successfully synthesized T-carbon nanowires using pseudo-topotactic conversion of carbon nanotubes under picosecond laser irradiation[17]. They found that T-carbon nanowires exhibit absorption in the ultraviolet region and strong photoluminescence around 436 nm, confirming their wide bandgap characteristics. Two years later, Kai et al.[16] synthesized T-carbon on polycrystalline and single-crystal diamond substrates using plasma-enhanced chemical vapor deposition.

In addition to research on bulk T-carbon and T-carbon nanowires, attention has also been given to the properties and applications of different T-carbon surfaces. In 2022, Guo et al.[18] used first-principles calculations to study the adsorption behavior of Na atoms on the T-carbon (111) surface, identifying the most favorable adsorption sites and demonstrating that the dielectric loss of the adsorption system decreased, which would benefit the longevity of electronic devices. In 2023, Zhao et al.[19] investigated the enhanced multifunctional electrocatalytic performance of T-carbon [110] monolayers as two-dimensional electrocatalyst substrates, modified/doped with transition metal and nonmetal atoms. They proved that such co-doped monolayers are excellent synergistic trifunctional electrocatalysts and that T-carbon monolayers have significant potential in electrocatalysis.

To retain the tetrahedral characteristics of T-carbon, preserving the out-of-plane carbon atoms on the T-carbon (111) surface creates dangling bonds. Surface dangling bonds or defects can significantly affect nanomaterials' optical and electronic properties[20]. Surface passivation is an effective approach to address these issues, reducing or even eliminating the impact of surface states and improving the stability of nanomaterials in atmospheric, aqueous, and thermal environments. Effective surface passivation can reduce or eliminate surface/defect states, creating a chemically inert surface that inhibits the formation of new surface states[20].

Exploring the intrinsic mechanisms of passivation is important due to its significant impact on the physical properties of materials. However, the experimental study of surface effects on nanomaterials is challenging due to the requirements of extremely clean surfaces, perfect detection conditions (e.g., ultra-high vacuum), and highly precise analytical instruments. Theoretical studies using first-principles calculations are an exploratory alternative to the more costly experimental approaches and have been extensively used in the literature to investigate these mechanisms[21-25].

Hydrogen is one of the most common passivating agents. Even in high-vacuum systems, $H_2O$ can exist as a residual gas and serve as a source of hydrogen groups. One-dimensional semiconductor nanomaterials are typically coated with oxide layers, which can be removed by high-frequency etching to provide hydrogenated surfaces. The effect of hydrogen passivation on the structural stability and electronic properties of Si[26], SiC[27, 28], ZnO[29], and ZnS[30] nanowires has been extensively investigated using first-principles calculations. Generally, hydrogen passivation enhances stability and widens the electronic bandgap compared to non-passivated materials, which can cause a transition from direct to indirect bandgap.

However, many other passivating agents are available besides hydrogen. On the surfaces of certain materials like Ge, hydrogen passivation layers are only stable for a few minutes under ambient environmental conditions[31]. Thus, other passivating agents, such as hydroxyl[23, 32, 33] and halogens (F, Cl, Br, I)[34-36], are considered. These are often present under experimental conditions involving solutions or hydrofluoric acid.

In this work, based on first-principles simulations, we have investigated the structural stability, optical and electronic properties of the T-carbon (111) surface passivated with different atoms and functional groups bonded to the unsaturated carbon atoms of the T-carbon tetrahedral tips. We also addressed the best passivating agents for different applications.

## 2. Calculation method and model

### 2.1 Calculation method

We have carried out first-principles calculations, density functional theory (DFT) level, using the CASTEP package[37] available in the Materials Studio (MS) 2020 suite to investigate the structural, optical, and electronic properties of T-carbon (111) passivated slabs. In particular, we investigated the changes induced by the structural passivation with different atoms and functional groups.

In all simulations, the exchange-correlation functional by the Perdew Burke Ernzerhof (PBE) version of the generalized gradient approximation (GGA) and the ultrasoft pseudopotential were employed to describe the interaction between the electrons and ions. Well-converged results are obtained using an energy cutoff of 400 eV and a k-point grid set of $4 \times 4 \times 1$. In order to eliminate any spurious effects created by the mirror image interactions, a standard vacuum buffer layer larger than 20 Å over the surface plane was used. The convergence quality was set to be ultrafine. The convergence tolerance of the maximum force acting on each atom during the relaxation and properties calculation processes was 0.01 eV/Å, and the total energy error was less than $5.0 \times 10^{-6}$ eV. In addition, the maximal stress and displacement were set to 0.02 GPa and $5.0 \times 10^{-4}$ Å, respectively. The $5.0 \times 10^{-7}$ electron convergence during the self-consistent field (SCF) calculations ensures high-quality results.

## 2.2 Calculation model

The selected structure of the present study was the (111) T-carbon crystal plane. This choice was based on the fact that this surface corresponds to the observed most intense diffraction peak and has the lowest surface energy[15].

The structural models were created in the following way. First, the bulk T-carbon was fully geometrically optimized (relaxed atomic positions and crystal lattice). The optimized lattice constant of optimized bulk T-carbon was 7.50 Å, consistent with previous theoretical and experimental results reported in the literature[15, 17, 18]. The phonon spectrum of the bulk T-carbon has no imaginary (negative) frequencies along the entire Brillouin zone (Fig. S1 (a)). Its electronic band structure (Fig. S1 (b)) is consistent with the reports calculated using the VASP package[15].

Then, a T-carbon slab is obtained by cleaving the relaxed bulk T-carbon along the [111] direction, followed by replicating the unit cell by 2×2 (Fig. 1 (a)). Furthermore, the T-carbon (111) surface is not relaxed, as shown in Fig. 1 (b). The carbon tetrahedron is distorted, and the T-carbon (111) surface cannot maintain its original atom arrangement and configuration, confirmed by C-C bond changes and virtual frequency phonon spectrum (Fig. S2).

Therefore, the pristine T-carbon (111) slab is then just used as a template for subsequent structural investigations. Our strategies for stabilizing T-carbon (111) are decorating the surface carbon atoms with different atoms and functional groups. As displayed in Fig. 1 (c), the top (blue balls) and bottom (green balls) layer carbon atoms in T-carbon (111) are terminated by different atoms and functional groups (pink balls), including -H, -O=, -OH, -NH$_2$ and -CH$_3$ (details see Fig. 1 (d)). For convenience, the -H, -O=, -OH, -NH$_2$ and -CH$_3$ passivated T-carbon (111) are denoted as -H, -O=, -OH, -NH$_2$ and -CH$_3$ decorated 2-(111) planar T-carbon.

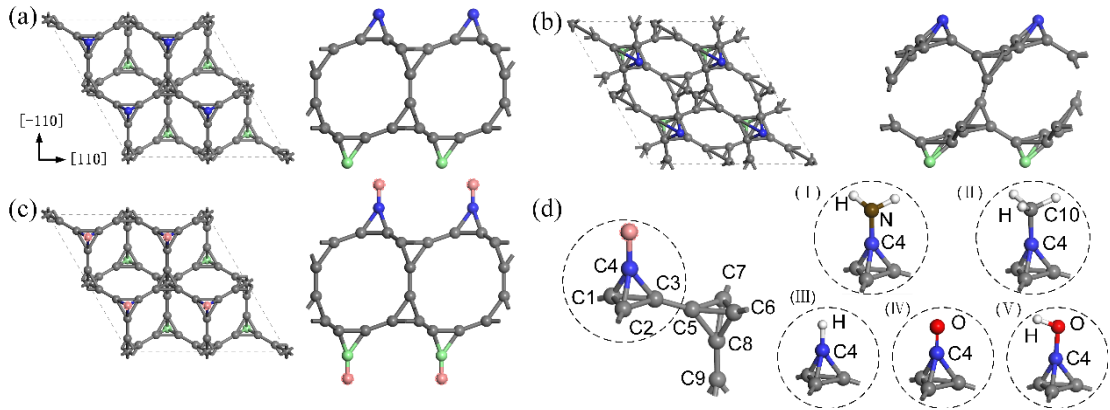

**Fig. 1** Views from [111] (left panel) and [110] (right panel) directions of (a) pristine, (b) optimized, and (c) surface terminated 2-(111) planar T-carbon structures. (d) the details of decorated functional groups and corresponding atom labeling. The functional groups include (I) -NH$_2$, (II) -CH$_3$, (III) -H, (IV) -O= and (V) -OH. The gray, blue, green, red, brown, white, and pink balls represent atoms in bulk, top surface, bottom surface carbon atoms, oxygen, nitrogen, hydrogen atoms, and decorated functional groups, respectively.

## 3. Results and discussion

### 3.1 Phonon spectrum for hydrogenated 2-(111) T-carbon

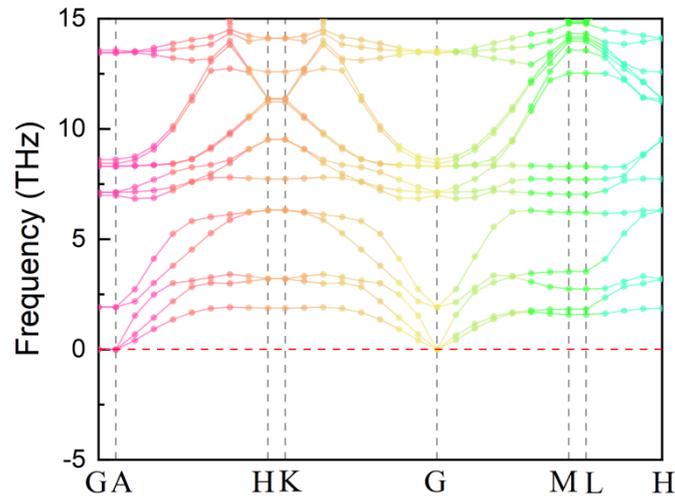

**Fig. 2** Phonon spectrum of H-decorated 2-(111) planar T-carbon.

Surface passivation presents a promising and effective strategy to improve the dynamic stability of 2-(111) planar T-carbon. To further confirm the stability of the present proposed systems, the phonon dispersion curves of -H decorated 2-(111) planar T-carbon through the whole Brillouin zone were calculated and presented in Fig. 2. Since there have no imaginary (negative) frequencies in the phonon dispersion spectra, we conclude that the dynamic stability of -H decorated 2-(111) planar T-carbon is verified, at least at T=0 K.

## 3.2 Electronic structure

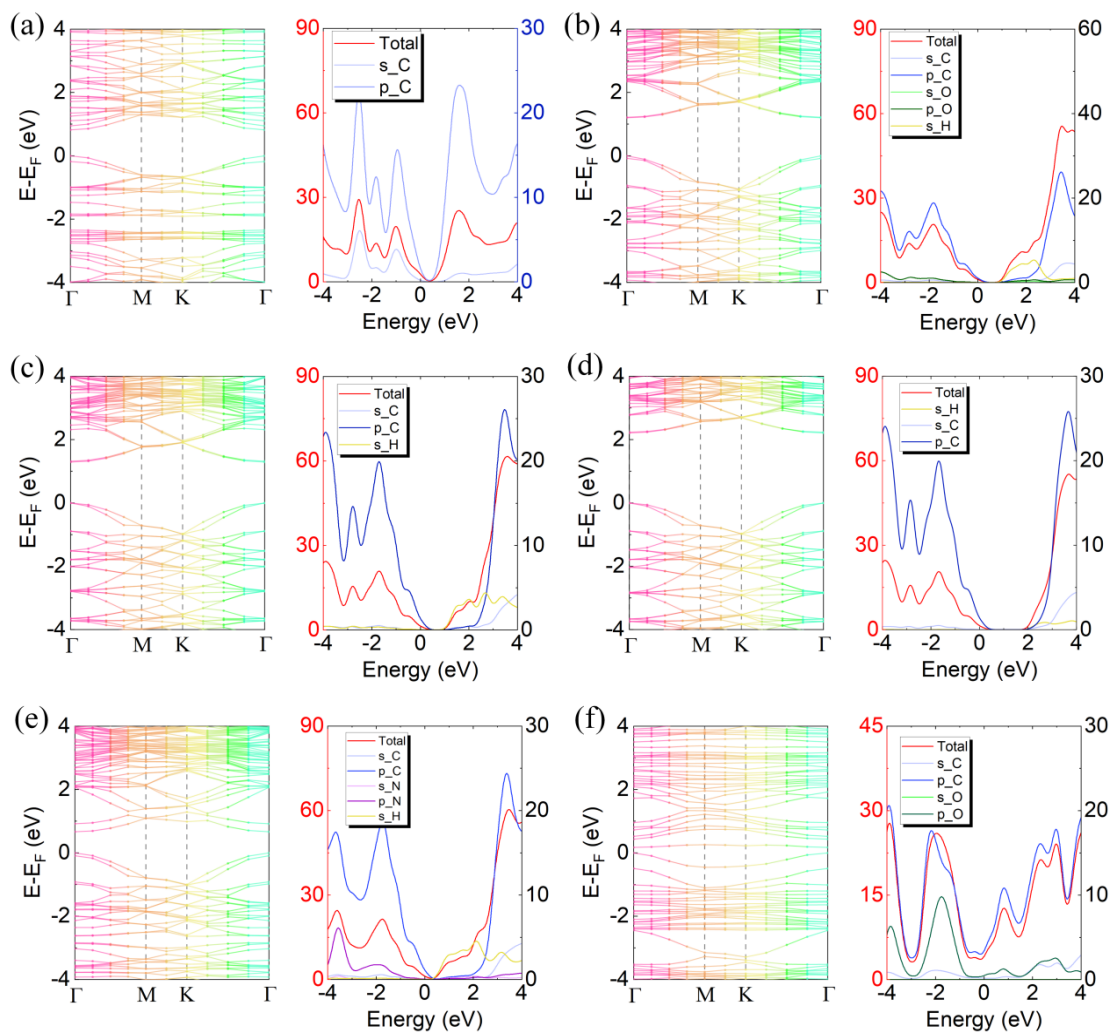

**Fig. 3** Electronic band structure and the corresponding density of states (DOS) of (a) pristine, (b) -OH, (c)-CH$_3$, (d)-H, (e)-NH$_2$ and (f) -O= decorated 2-(111) planar T-carbon.

In **Fig. 3**, we present the electronic band structure (left panel) and the corresponding density of states (DOS) (right panel) of pristine, -OH, -CH$_3$, -H, -NH$_2$, and -O= decorated 2-(111) planar T-carbon. Herein, the energy level of 0 eV is defined as Fermi energy. It can be seen from **Fig. 3** that for the pristine, -OH, -CH$_3$, -H, -NH$_2$, and -O= passivated 2-(111) planar T-carbons, all of them have direct gap bonds at the Γ point. The corresponding band gap values are 0.82, 1.22, 1.29, 2.21, 0.67, and 0.17 eV, respectively, typical values of semiconductors.

The gap values increase gradually from pristine to -OH, -CH3, and -H. However, the values decrease gradually from pristine to -NH$_2$ and -O=, suggesting the band gap of 2-(111) planar T-carbon can be tuned at a large range from narrow to wide band gaps by surface decorations. Besides proving the possibility of tuning the electronic behavior for a large energy range, the passivation also improves the structural stability of 2-(111) planar T-carbon.

In order to further understand the features of the electronic band structure, we have also calculated the total density of states (TDOS) and projected density of states (PDOS) of the structures mentioned above, as shown in the right panel of **Fig. 3**. For -OH, -CH$_3$, -H, and -NH$_2$, in the energy interval of the PDOS from -4.0 to 0 eV, which is the valence band regions, the major contributions came from the p_C orbitals. The major contributions in the conduction band regions are also from the p_C orbitals. In contrast, the contributions from s_C and s_H are small. For the -O= passivated structure, the PDOS main contributions came from p_C and p_O orbitals below the Fermi level. For the conduction bands above the Fermi level, the PDOS is mainly formed by p_C orbitals and small contributions from the p_O orbitals.

## 3.3 Bond length and bond population

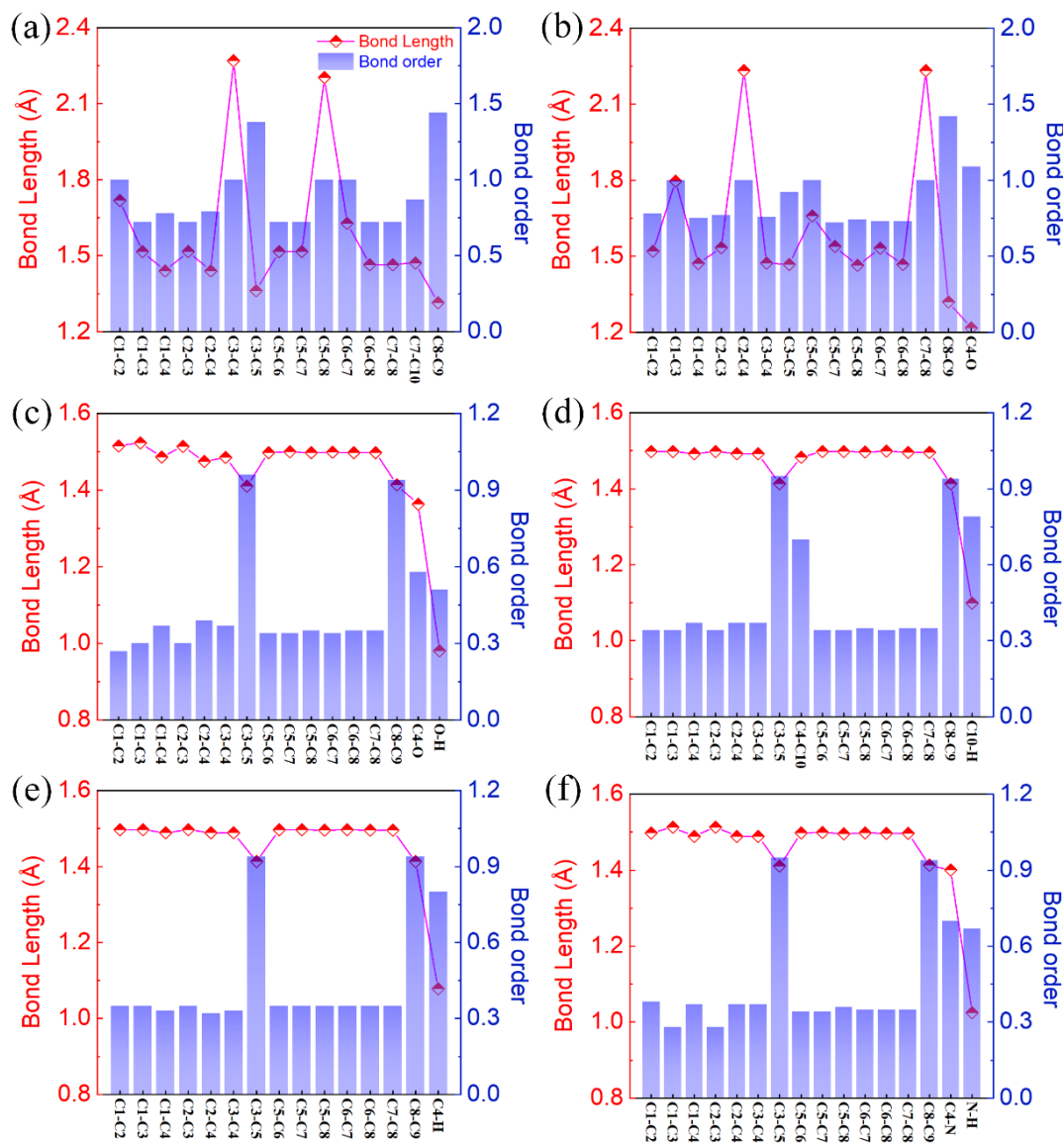

**Fig. 4.** Bond length and population of (a) pristine, (b)-O=, (c)-OH, (d)-$CH_3$, (e)-H and (f)-$NH_2$ decorated 2-(111) planar T-carbon.

In **Fig. 4**, we present the selected atom distances and bond order values of pristine and different passivated structures. The atom distances distribution of pristine 2-(111) planar T-carbon is large, with some distances larger than 2.2 Å (~1.417 Å is the typical distance in bulk T-carbon), indicating the atoms are not bonded anymore. We predict that it is because of the instability of the 2-(111) planar T-carbon surface and much stronger inter-tetrahedral bond than intra-tetrahedral one, both inducing bond distortions. Similar behavior was also seen in -O= decorated 2-(111) planar T-carbon, which is due to the unbalanced tetrahedral carbon atoms by surrounding atoms with significant differences in bond strength. Compared to the pristine 2-(111) planar T-carbon, the lengths of inter-tetrahedral bonds, namely C3-C5 and C8-C9 in the functional group decorated systems, except for C8-C9 bond in -O= structure, are increased and closer to that (~1.417 Å) of bulk T-carbon and cleaved non-passivated 2-(111) planar T-carbon. The variation of intra-tetrahedral bond length is significantly decreased, indicating that the decorated 2-(111) planar T-carbon can maintain its initial atom arrangement and configuration without significant structural distortions. The bond population presents a uniformly inverse correlation with the bond length values. Compared to the inner bonds, the bond length formed between the surface C atom, i.e., C4, and the passivated atom is relatively short. By analyzing the bond lengths and populations, one can conclude that the surface passivation of 2-(111) planar T-carbon by different functional groups can distinctly enhance the overall stability of the non-passivated 2-(111) planar T-carbon.

## 3.4 Optical properties

### 3.4.1 dielectric functions

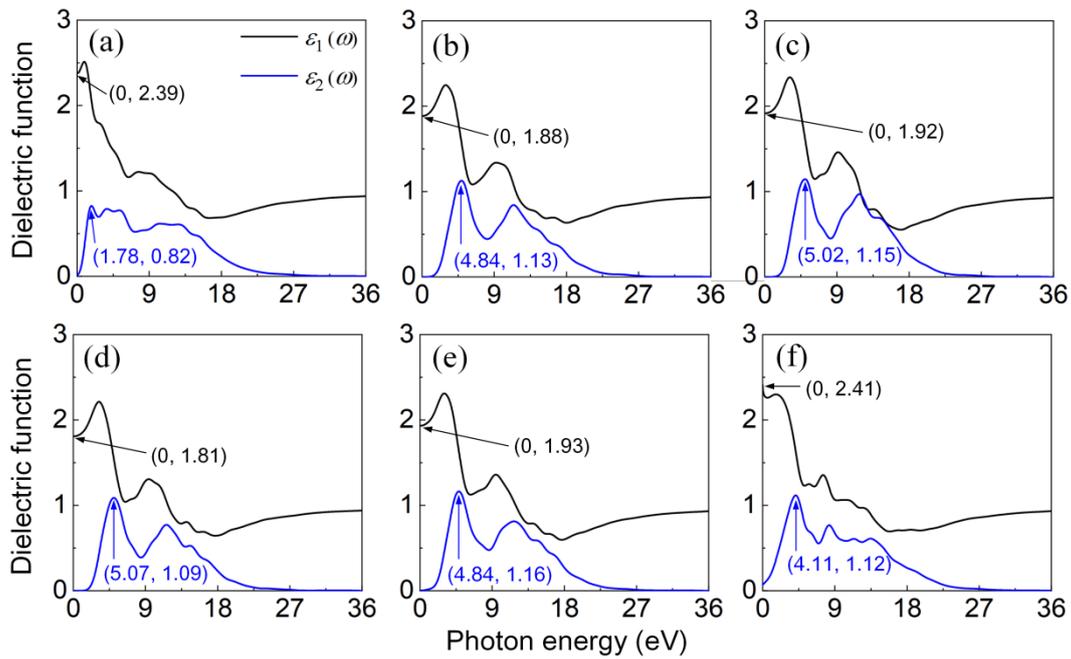

**Fig. 5** Dielectric function as a function of the photon energy of (a) pristine, (b) -OH, (c)-CH$_3$, (d)-H, (e)-NH$_2$, and (f) -O= decorated 2-(111) planar T-carbon. Here, $\varepsilon_1(\omega)$ and $\varepsilon_2(\omega)$ are the real and imaginary parts of the dielectric function, respectively. $\omega$ is the energy frequency.

When performing calculations of optical properties of nanostructures due to electronic transitions,

it is usual to evaluate the complex dielectric constant and then express other properties in terms of it to describe the optical response of the materials. The complex dielectric function, $\varepsilon(\omega)$, is calculated by $\varepsilon(\omega)=\varepsilon_1(\omega)+i\varepsilon_2(\omega)$, where, $\varepsilon_1(\omega)$ is the real part of the dielectric function representing the storage capacity of electromagnetic energy, and $\varepsilon_2(\omega)$ is the imaginary part of the dielectric function characterizing the loss of electromagnetic energy.

In **Fig. 5**, we present the dielectric function of pristine and different functional groups decorated 2-(111) planar T-carbon as a function of the photon energy. The black and light blue curves represent the real part $\varepsilon_1(\omega)$ and imaginary part $\varepsilon_2(\omega)$, respectively. From the pristine, -OH, -CH$_3$, -H, -NH$_2$ to -O= decorated 2-(111) planar T-carbon, the overall trend of the real part decreases gradually with the increase of photon energy. The static dielectric constant is 2.39, 1.88, 1.92, 1.81, 1.93, and 2.41, respectively. Compared with the pristine one, the static dielectric constants of decorated 2-(111) planar T-carbon all decreased except for the -O= case, which has almost the same value.

Materials with low dielectric constant usually have low electric fields and dielectric losses, making them suitable for high-frequency electronics and microwave communication systems. The main peaks, related to dispersion, in the spectra of the real part appear at 0.90, 2.99, 3.11, 3.21, 3.09, and 1.59 eV in sequence. The static dielectric constant of all passivation systems increased to a certain extent, in which the -CH$_3$ decorated structure is the most evident. Moreover, the imaginary part of the dielectric constant in pristine 2-(111) planar T-carbon reaches the maximum value of 0.82, located at 1.78 eV in the low-energy region, in which the dielectric loss reaches the maximum. Its imaginary part of the dielectric constant exhibits an overall downward trend with the increase of phonon energy, accompanied by some peaks in the energy range from 2.67 eV to 35.0 eV, which derives from the interband transitions between the valence and conduction states. Finally, the imaginary part approaches zero as the energy exceeds 35.0 eV, manifesting no dielectric loss.

The imaginary part of the dielectric constant in decorated 2-(111) planar T-carbon illustrates changes similar to those of pristine structures. As to -OH, -CH$_3$, -H, -NH$_2$ to -O= decorated 2-(111) planar T-carbon, the first peaks are 1.13, 1.15, 1.09, 1.16, and 1.12 appearing at 4.84, 5.02, 5.07, 4.85 and 4.11 eV, respectively, therewith followed by some weak peaks arising from interband transitions. Compared to the pristine 2-(111) planar T-carbon, the maximum values of the imaginary part of the dielectric constant significantly uniformly increase and move to the high energy region (shorter wavelength region). These results confirm that the blue shift phenomenon occurs in the passivated 2-(111) planar T-carbon. The dielectric constant and dielectric loss can be effectively modulated by surface passivation, providing a guideline for optical property tuning and widening application scenarios of T-carbon slab in the field of dielectric materials.

### 3.4.2 Absorption and reflectivity spectra

The absorption coefficient characterizes the light intensity decreases with the propagation distance or penetration depth when the light propagates in the medium. In **Fig. 6**, we present the simulated absorption spectra for pristine and functional groups decorated 2-(111) planar T-carbon. The change trends of all absorption spectra are basically the same as the phonon energy increase, and three main absorption peaks with different intensities are present; finally, the curves rapidly approach 0 due to electronic transitions. The curves reach a maximum value of 6.46, 7.42, 8.73, 7.05, 7.67, and 6.97 ($\times 10^4$ cm$^{-1}$) located at 13.37, 12.07, 12.24, 12.33, 12.74, and 14.40 eV, respectively. Compared with the pristine one, other peak values increase and shift to the lower energy region except for -O= decoration, indicating a redshift appearance. These modified 2-(111) planar T-carbon with a large direct band gap can be potential candidates for photovoltaic applications possessing large adsorption and small reflectivity in the visible light range.

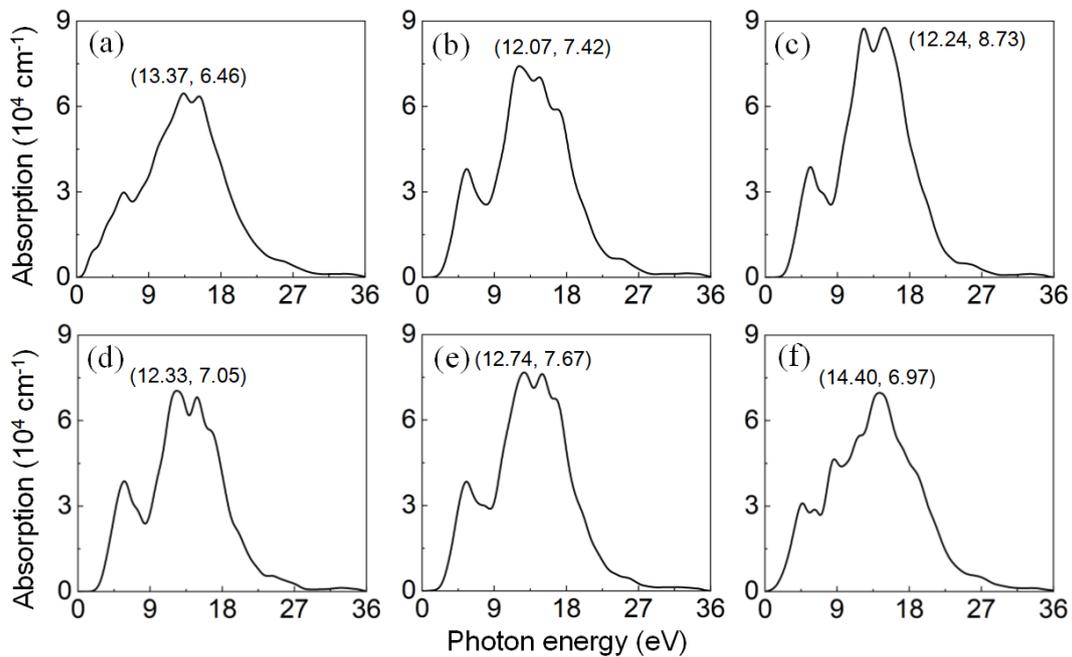

**Fig. 6.** Simulated absorption spectra as a function of the photon energy of (a) pristine, (b) -OH, (c)-CH$_3$, (d)-H, (e)-NH$_2$, and (f) -O= decorated 2-(111) planar T-carbon. The peak value of each curve and corresponding photon energy are marked in the figures.

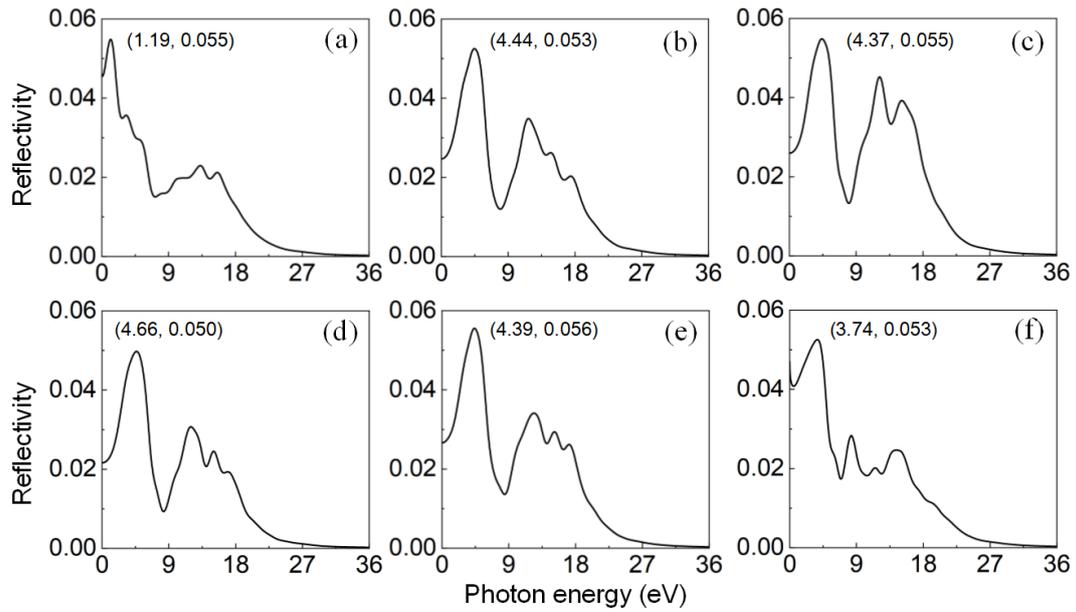

**Fig. 7.** Simulated reflection spectra as a function of the photon energy of (a) pristine, (b) -OH, (c)-CH$_3$, (d)-H, (e)-NH$_2$, and (f) -O= decorated 2-(111) planar T-carbon. The maximum value of each curve and corresponding photon energy are marked in the figures.

The reflection spectrum represents the capacity to reflect electromagnetic radiation when light is directly and vertically incident from air to the medium surface. The reflection spectra for pristine and functional groups decorated 2-(111) planar T-carbon are presented in **Fig. 7**. With the increase of photon energy, the reflection spectra exhibit two prominent peaks, then gradually decrease close to 0. For pristine, -OH, -CH$_3$, -H, -NH$_2$, and -O= decorated 2-(111) planar T-carbon, the maximum values of reflection intensity are 0.055, 0.053, 0.055, 0.050, 0.056, and 0.053 located at 1.19, 4.44, 4.37, 4.66, 4.39, and 3.74 eV, respectively. Compared with the pristine one, the peak values are roughly equivalent and have shifted to the higher energy region, exhibiting a blueshift.

### 3.4.3 Loss function

The loss function describes the energy loss of the electrons passing through a uniform dielectric medium. The loss functions for pristine and functional group decorated 2-(111) planar T-carbon are presented in **Fig. 8**. With the increase of photon energy, for pristine, -OH, -CH$_3$, -H, -NH$_2$, and -O= decorated 2-(111) planar T-carbon, the loss spectra reach the maximum values of 0.63, 0.68, 0.87, 0.66, 0.73, and 0.65 at phonon energy of 15.73, 15.18, 16.45, 15.29, 17.18, and 15.36 eV, respectively. The loss functions approach 0 when the photon energy exceeds 35.0 eV. Compared with the pristine one, other peak values increase. The peaks move toward higher energy values for -CH3 and -NH2 decorated 2-(111) planar T-carbon, resulting in an energy blueshift. On the contrary, the maximum loss function values for -OH, -H, and -O= decorated 2-(111) planar T-carbon shift to the lower energy values, indicating a redshift.

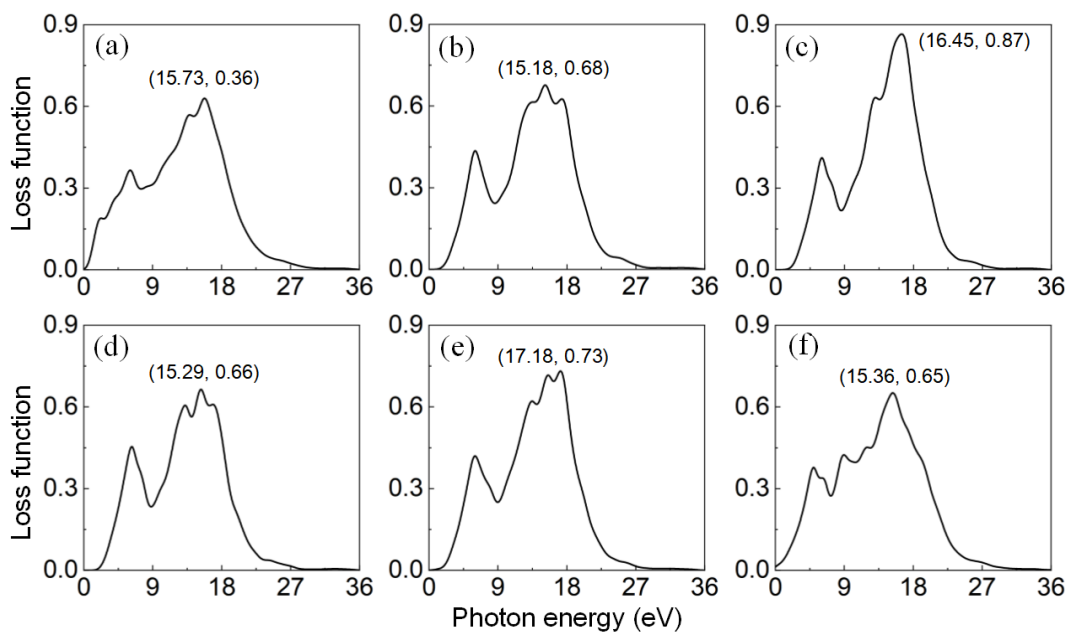

**Fig. 8.** Loss function of (a) pristine, (b) -OH, (c)-CH$_3$, (d)-H, (e)-NH$_2$ and (f) -O= decorated 2-(111) planar T-carbon as a function of the photon energy. The peak value of each curve and corresponding photon energy are marked in the figures.

## 4. Conclusions

In this work, we proposed a novel two-dimensional material, named 2-(111) planar T-carbon, and investigated its structural stability, electronic, and optical properties of pristine and functionalized by five different groups (-OH, -CH$_3$, -H, -NH$_2$, and -O=) through DFT simulations. The main conclusions are given below:

(1) The surface passivation of 2-(111) planar T-carbon by different functional groups can significantly decrease the variation of intra-tetrahedral bond length and, therefore, significantly enhancing the overall structural stability of the non-passivated 2-(111) planar T-carbon.

(2) The pristine and five passivated 2-(111) planar T-carbons all have direct electronic band gaps at the Γ point. The band gap of 2-(111) planar T-carbon can be tuned from 0.17 eV (-O=) up to 2.21 eV (-H) by specific functional group passivation.

(3) The imaginary part of the dielectric constants and reflection spectra uniformly increase significantly from 1.78 eV to 5.07 eV, and 1.19 eV to 4.66 eV, respectively, confirming energy blue shifts for all the passivated structures, with the exception of the -O= case, where a red shift for absorption curves from 13.37 eV to 12.07 eV was observed.

In summary, this work proposes an easy way to improve the structural stability of 2-(111) planar T-carbon and shows the possible tuning of electronic and optical properties of 2-(111) planar T-carbon through selective passivation. This information can be helpful for the design of potential T-carbon-based photoelectric devices.

## Supplementary materials

Supporting information.doc

## Funding

The authors gratefully acknowledge the financial support from Development and Reform Commission of Shenzhen (No. XMHT20220103004), the Natural Science Foundation of Guangdong (2024A1515010821) and the Nature Science Research Project of Yan'an University (YDBK2022-89). DSG acknowledges financial support from the Center for Computational Engineering and Sciences at Unicamp through the FAPESP/CEPID (No. 2013/08293-7).

## Author Contributions

H. Cai and Z. Duan prepared the model, collected data, prepared the figures; H. Cai and K. Cai prepared the manuscript; K. Cai proposed the idea and managed the project; D.S. Galvao reviewed the manuscript. All authors attended in discussion in preparing the manuscript.

## Interest conflict

The authors declare that they have no conflict of interest.